\documentclass[twocolumn,showpacs,preprintnumbers,amsmath,amssymb]{revtex4}
\usepackage{bm}
\usepackage{epsfig}
\usepackage{graphicx}
\usepackage{dcolumn}
\usepackage{amsfonts}

\begin{document}

\title{Asymmetry in the reconstructed deceleration parameter}

\author{Carla Bernal}
\email{carlabernalb[at]gmail.cl}

\author{V\'ictor H. C\'ardenas}
\email{victor.cardenas[at]uv.cl}

\author{Veronica Motta}
\email{veronica.motta[at]uv.cl}

\affiliation{Instituto de F\'{\i}sica y Astronom\'ia, Facultad de
Ciencias, Universidad de Valpara\'iso, Av. Gran Breta\~na 1111,
Valpara\'iso, Chile}

\begin{abstract}
We study the orientation dependence of the reconstructed
deceleration parameter as a function of redshift. We use the Union 2
and Loss datasets, by using the well known preferred axis discussed
in the literature, finding the best fit reconstructed deceleration
parameter. We found that a low redshift transition of the
reconstructed $q(z)$ is clearly absent in one direction and
amazingly sharp in the opposite one. We discuss the possibility that
such a behavior can be associated with large scale structures
affecting the data.
\end{abstract}

\maketitle

\section{Introduction}

Although the $\Lambda$CDM model still remains as the simplest model
that fits the observational data, implying the universe is currently
expanding at an accelerated rate \cite{snia1, snia2} driven by the
cosmological constant $\Lambda$, the model is unable to explain both
the order of magnitude of its strength and why we live in such a
special epoch where both the contribution of this constant and of
dark matter are of the same order of magnitude.

As a departure from this cosmological constant contribution,
cosmologists have proposed the idea of dark energy (DE)
\cite{dereview}, a theoretical fluid model characterized by a
dynamical equation of state (EoS) parameter $p/\rho=w(z)$, with $z$
being the redshift. If this parameter is constant and equal to
$w=-1$, it corresponds to the cosmological constant. The source of
this dynamical DE could be either a new field component filling the
universe, as a quintessence scalar field \cite{quinta}, or a
modifying gravity \cite{modgrav}.

Another venue explored to tackle this problem is based on relaxing
the Copernican Principle (CP). This was, in fact, proposed shortly
after the discovery of the accelerated expansion
\cite{pascualsanchez},\cite{celerier00}. In this case, the effect of
DE could be explained away by large-scale nonlinear inhomogeneities.
Almost all the work have been based on the spherically symmetric
Lemaitre-Tolman-Bondi (LTB) model. In this context, a typical
scenario that can mimic the late-time behavior of the $\Lambda$CDM
model consists of positioning the observer near the center of a deep
void $1-3$ Gpc wide. The vast literature on inhomogeneous studies,
in particular the LTB models, is summarized in \cite{bolejko11}, and
\cite{marranotari11}.

Although it seems unnatural to assume we live inside a huge void
(larger than the observed ones), the lesson here is that by
modifying the theory through which we interpret the data, it is
possible to abandon the also unnatural assumption of the existence
of a negative pressure fluid acting only recently.

Assuming the CP is valid, direct explorations of possible departures
from the $\Lambda$CDM model have been focused mainly through
parameterizing specific physical quantities. For example, there are
models where the EoS parameter $w(z)$ is parameterized, others where
the deceleration parameter $q(z)=-\ddot{a}a^2/\dot{a}$ is
parameterized, and also those where the DE density $\rho_{de}(z)$ is
directly parameterized.

One of the lessons we have learned from these studies is the
apparent existence of a tension between the low-redshift
observational constraints, and those from high-redshift. For
example, there is evidence that the constraints obtained from SNIa,
or gas mass fraction in galaxy clusters, points to a different
region in the parameter space than those derived from baryon
acoustic oscillation (BAO) and cosmic microwave background (CMB)
data \cite{shafi2009,Li:2010da,
Cardenas:2011a,Cardenas:2013roa,cardenas:2014,Magana:2014voa}

As was showed in \cite{cardenas:2014}, once we use models that
allows variations, whether in the EoS parameter $w(z)$ or in the DE
density $X(z)=\rho_{de}(z)/\rho_{de}(0)$, the reconstructed
deceleration parameter $q(z)$ shows a low-redshift transition when
SNIa and gas mass fraction in cluster datasets are used. The
evidence for such transition disappears once data from BAO and CMB
are added. Then an important connection was recognized. In
\cite{Vanderveld:2006rb, february2010} the authors showed that,
using the data from SNIa through the luminosity distance obtained
from an LTB model, a {\it similar} low-redshift transition in the
effective deceleration parameter appears, implying that such a
feature can be considered a signal of the existence of a void
(underdensity) in our neighborhood.

A similar tension between the low-redshift observational constraints
and those from high-redshift also manifest in measurements of $H_0$,
as was recently studied in \cite{Verde:2013wza}, where a strong
tension is found by comparing local measurements of $H_0$ and $t_0$
with those from Planck. This tension emerges using the $\Lambda$CDM
model as the ``base'' model. Although simple extensions of the
$\Lambda$CDM model were studied, not all of them alleviated the
tension. In \cite{Keenan:2013mfa} the authors showed that the
assumption of a large local under-density on radial scales of a few
Mpc it is enough to alleviate such a tension.

Besides the already mentioned tension, during the last years several
works indicate the existence of a preferred direction of expansion.
For example, the large scale bulk-flow
\cite{Watkins:2008hf,Kashlinsky:2008us,Kashlinsky:2008ut,Feindt:2013pma,Atrio-Barandela:2014nda},
the low multipoles alignment in the CMB angular power spectrum
\cite{Akrami:2014eta}, the large scale alignment of quasar
polarization \cite{hutsemekers}, and studies using SNIa
\cite{antoniou,Dai:2011xm,Colin:2010ds,cai_tuo,Mathews:2014fma,Jimenez:2014jma,Appleby:2014lra,Appleby:2014kea,Bengaly:2015dza}.
These studies reported a dipole anisotropy within $1-2 \sigma$
confidence level intriguingly pointing towards the same direction.
It is also interesting to notice that some authors conclude that
such asymmetry in SNIa data could emerge from a systematic error
\cite{javanmardi15} or simply it does not exist
\cite{Huterer:2015gpa}. Of course, there is also the interesting
possibility that this behavior is caused by new physics.

Certainly both, a possible existence of a huge structure in our
neighborhood, and the asymmetry in parameter fitting emerging from
the use of low-redshift data, are connected. In this way we use the
result of \cite{Vanderveld:2006rb} as our main hypothesis in this
work.

In this paper we want to use the reconstructed deceleration
parameter as a probe for hints of a local under-density and then
delve into the effects of orientation on the reconstruction of the
deceleration parameter.

It is the purpose of this paper to explore how the reconstructed
deceleration parameter behaves once the position of the SNIa in the
sky is taken into account. In the next section we discuss the data
we use in the calculations, and present the procedure used to find
the change in the deceleration parameter. In section III we discuss
the orientation dependence separating the data in hemispheres. Next,
we describe our analysis using four regions with equal number of
data points. Finally, we show and discuss the results.

\section{The Union 2 data set}

In our previous work on the reconstruction of the deceleration
parameter we have made use of different SNIa datasets. For example,
we have worked with the Constitution \cite{constitution}, the Union
2 \cite{Union2}, the Union 2.1 \cite{suzuki}, and the Loss-Union
\cite{Ganeshalingam:2013mia} sets. Because we are interested in
studying the effects of the orientation dependence of the data we
decided to use a well studied dataset for which different methods
have found the same dipole anisotropy.
In this work we chose to work with the Union 2 set.

The Union 2 dataset consist of 557 SNIa in the range $0.015<z<1.4$.
Union 2 join the Union dataset \cite{Union}, six SNIa at high
redshift found by the Hubble Space Telescope, and low and
intermediate redshift data from \cite{hicken} and \cite{holtzman},
respectively.

To start with, let us perform the reconstruction using all the data,
without considering their position in the sky. This is actually a
calculation already made in \cite{Li:2010da} and
\cite{Cardenas:2011a}.

\begin{figure}[h!]
\centering \leavevmode\epsfysize=4.5cm \epsfbox{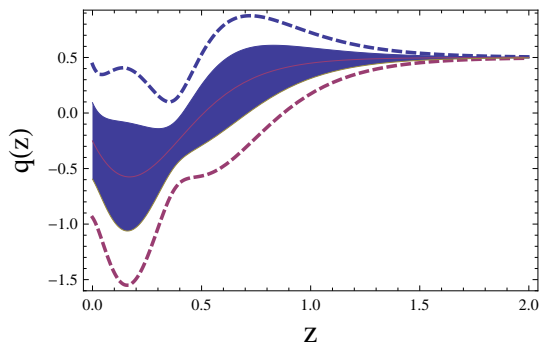}\\
\caption{Using the Union2 data set we plot the deceleration
parameter with error propagation at $2\sigma$. Here is possible to
note a transition of $q(z)$ a low-redshift at $1\sigma$
C.L.}\label{fig:fig1}
\end{figure}

We begin recalling that the co-moving radial coordinate is
\begin{equation}\label{comdistance}
r(z) =  \frac{c}{H_0} \int_0^z \frac{dz'}{E(z')},
\end{equation}
in a flat universe, where $E(z)=H(z)/H_0$ is the reduced Hubble
function. Assuming an EoS parameterized by
\cite{Chevallier:2000qy, Linder:2003nc}
\begin{equation}\label{cpl}
w(z)=w_0 + \frac{w_1 z}{1+z},
\end{equation}
the reduced Hubble function takes the form
\begin{equation}\label{edez}
E^2(z)  = \Omega_m (1+z)^3 +
(1-\Omega_m)X(z),
\end{equation}
where $\Omega_{m}$ comprises both the baryonic and non baryonic DM,
and \begin{equation}\label{cplX} X(z)= e^{-\frac{3 w_1 z}{1+z}}
(1+z)^{3 (1 + w_0 + w_1)}.
\end{equation}
From this, following previous works \cite{Li:2010da,
Cardenas:2011a}, we reconstruct the deceleration parameter function
\begin{equation}\label{qdzeq}
q(z) = (1+z)\frac{1}{E(z)}\frac{dE(z)}{dz}-1.
\end{equation}
Here we perform the calculation considering error propagation, and
display it in Figure \ref{fig:fig1}.
Using the best fit values found for the parameters -- $\Omega_m =
0.420 \pm 0.068$, $ w_0 = -0.86 \pm 0.38$, and $w_1 = -5.5 \pm 5.9$
-- the reconstructed deceleration parameter shows a rapid variation
at small redshift, a feature that is also apparent at $\sim 1
\sigma$ C.L.

Although intriguing, the feature has a low statistical significance.
To study the consistency of the data with the $\Lambda$CDM model we
follow the method proposed in \cite{perivola2010}, where the
distance $d_{\sigma}$ (in units of $\sigma$) from the best fit point
to the $\Lambda$CDM model is defined through the relation
\begin{equation}\label{dsigma}
1 - \Gamma(1,\Delta \chi^2/2)/\Gamma(1) =
\verb"Erf"(d_{\sigma}/\sqrt{2}),
\end{equation}
where the left hand side is the cumulative distribution function
(for two parameters), and $\Delta \chi^2 =
\chi^2_{(w_0,w_1)}-\chi^2_{min}$ is the $\chi^2$ difference between
the best fit and the $\Lambda$CDM point ($w_0=-1,w_1=0$). In our
case, the Union 2 set -- under this model -- is only $0.62 \sigma$
away from the $\Lambda$CDM.

As was shown in \cite{cardenas:2014}, the low redshift transition of
$q(z)$ at $2 \sigma$ C.L is observed in both the more recent SNIa
data and gas mass fraction data from galaxy clusters. The idea here
is to study how this picture changes once the position of the SNIa
data is taken into account.

\section{Testing the orientation dependence using hemispheres}

\begin{figure}[h!]
\centering \leavevmode\epsfysize=4.5cm \epsfbox{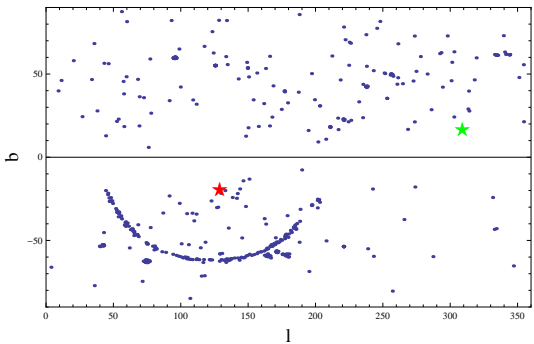}\\
\caption{Union 2 data set is plot using galactic coordinates. The
points marked by a green (red) star correspond to the points of
maximum (minimum) acceleration found in \cite{antoniou}.}
\label{fig2}
\end{figure}

In this section we make use of previous works using the Union 2 set,
which determine the maximum and minimum acceleration direction. In
particular, we use the result of \cite{antoniou} using the
hemispherical asymmetry technique, were they found a dipolar
asymmetry with a maximum acceleration towards $(l,b)=(309^o,18^o)$
and a minimum acceleration towards $(l,b)=(129^o,-18^o)$, in
galactic coordinates (see Fig.(\ref{fig2})).

We start, in this section, by separating the SNIa data in two sets,
each one belonging to an hemisphere. In the hemisphere corresponding
to the maximum acceleration there are 116 supernovae and in the
other the remaining 441. This huge asymmetry in the distribution is
essentially due to the fact that in the minimum acceleration
hemisphere, we found the SDSS stripe of SNIa data.

The best fit values in each hemisphere are displayed in Table
\ref{tab:table01}:

\begin{table}[h!]
\begin{center}
\begin{tabular}{ccccc}
\hline
\\
 & $\chi^2_{min}/dof$ & $\Omega_{m}$ & $\omega_0$ & $\omega_1 $ \\

\hline
A   &   $103.65/113$ & $0.21 \pm 0.64 $ & $ -1.3 \pm 1.2 $ & $1.5 \pm 3.4$  \\
B   &   $427.57/438$ & $0.46 \pm 0.06$ & $-0.41 \pm 0.53 $ &  $-9.2 \pm 7.3$\\
\hline
\end{tabular}
\end{center}
\caption{Best fit values of the cosmological parameters for the two
hemispheres. Row A for the 116 SNIa in the maximum acceleration
hemisphere. Row B the 441 SNIa in the opposite direction. See also
Fig.(\ref{fig00}). \label{tab:table01}}
\end{table}

\begin{figure}[h!]
\centering \leavevmode\epsfysize=9.0cm \epsfbox{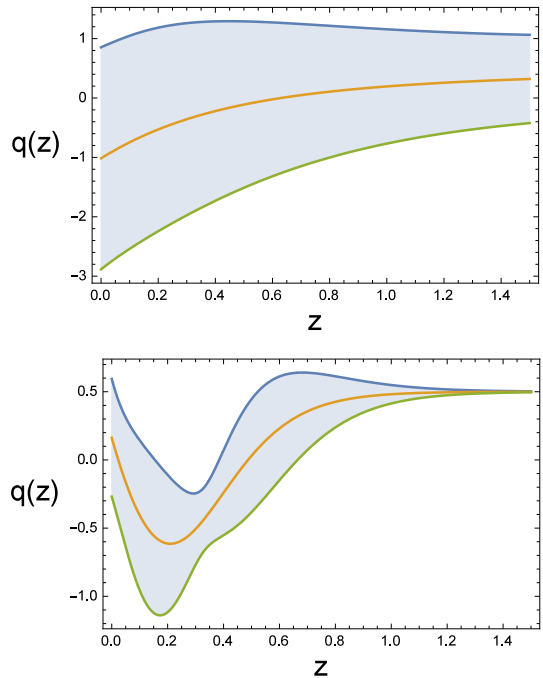}
\caption{ Here we show the reconstructed deceleration parameter
$q(z)$ using data from two hemispheres. The upper panel showing the
one around the maximum acceleration point, and the bottom the one in
the opposite direction. See Table \ref{tab:table01} for the best fit
values for each case. } \label{fig00}
\end{figure}

We notice that the transition feature at low redshift is only
evident in the direction of the minimum acceleration, meanwhile in
the opposite direction the best fit is in agreement with
$\Lambda$CDM. Following the same procedure as in the previous
section, we can estimate the confidence for each set of best fit
parameters. In the case of the 116 SNIa in the maximum acceleration
hemisphere we get a distance of $0.31\sigma$ away from the
$\Lambda$CDM point. In the other hemisphere -- in which we observe
the low redshift transition of the reconstructed $q(z)$ -- we get a
distance of $1.76\sigma$ away from the $\Lambda$CDM best fit.

We also notice a large difference in $\Omega_m$ best fit between
hemispheres. This shows us how much the model (our CPL $w_0w_1$CDM)
needs to stretch itself, once the direction of the SNIa is taken
into account. Using the result of \cite{Vanderveld:2006rb} as our
main hypothesis -- that a low redshift transition of $q(z)$ can be
considered a signal for a local under-density -- the result we have
found indicates the data (Union 2 set) is able to detect such an
under-density only in one (hemisphere) direction of the sky. This
implies that the assumption of a spherically symmetric model for
such under-density is not a good idea. Moreover, this result
probably suggest the use of a metric with a dipolar asymmetry.

At this point, we would like to discuss some of the preliminary
results we have found. First of all, we have to stress that only
SNIa data have been used. The low redshift transition apparent in
Figures \ref{fig:fig1} and \ref{fig00} disappear once data from
baryon acoustic oscillation (BAO) and cosmic microwave background
radiation (CMBR) are taken into account. In this sense, we are
exploring a possible cause of the original tension between low and
high redshift observational constraints discussed in the
introduction. In this case, as is also the case with a similar
tension in the determination of $H_0$, the origin seems to be, as
the authors in \cite{Keenan:2013mfa} suggest, the existence of a
local under density. However, in this work we add that such under
density is only effective in one direction on the sky.

It is also worth notice the effect of the number of points used in
each hemisphere. The set pointing towards the maximum acceleration
has only 116 SNIa. Although the best fit is only $0.3 \sigma$ away
from the $\Lambda$CDM values, indicating these data points are in
agreement with the concordance model, the sparsity of data points
increases the errors considerably. The actual redshift distribution
is also worth mentioned in this context. In this hemisphere most of
the data, around 65 points, have $z<0.2$, with no data in the range
$0.2<z<0.3$, with 45 points distributed around redshift $z \simeq
0.5$, and almost one point for every redshift bin ($\triangle z =
0.1$) for $z>0.8$. The other hemisphere has a greater number of data
points (441) with a redshift distribution that, although decreasing
with redshift, it still maintain a large number of data points per
redshift bin ($\triangle z = 0.1$) until $z \simeq 1.4$.

\section{Analysis by regions}

The results obtained in the previous section used two sets of data
with a very different number of points each. In this section we
select regions in the sky, around the points of maximum and minimum
acceleration, trying to get two sets of data of approximately the
same number of points.

Given that by separating in hemispheres produces that one of them
has only a hundred data points, we decided to study regions
containing at least 50 data as a minimum. Then, we increase the
number of points to 100, 150 and 200.

The regions consist of circular patches around the centers of
maximum and minimum acceleration (found through the hemisphere
comparison method by \cite{antoniou}). Given the anisotropic
distribution of data, the regions have very different sizes. In what
follow we present the analysis considering four regions: A with 50,
B with 100, C with 150 and D with 200 data points.

\subsection{Region A}

Let us discuss first the case having approximately 50 data points
around each region.

\begin{figure}[h!]
\begin{center}$
\begin{array}{c}
\includegraphics[width=2.0in]{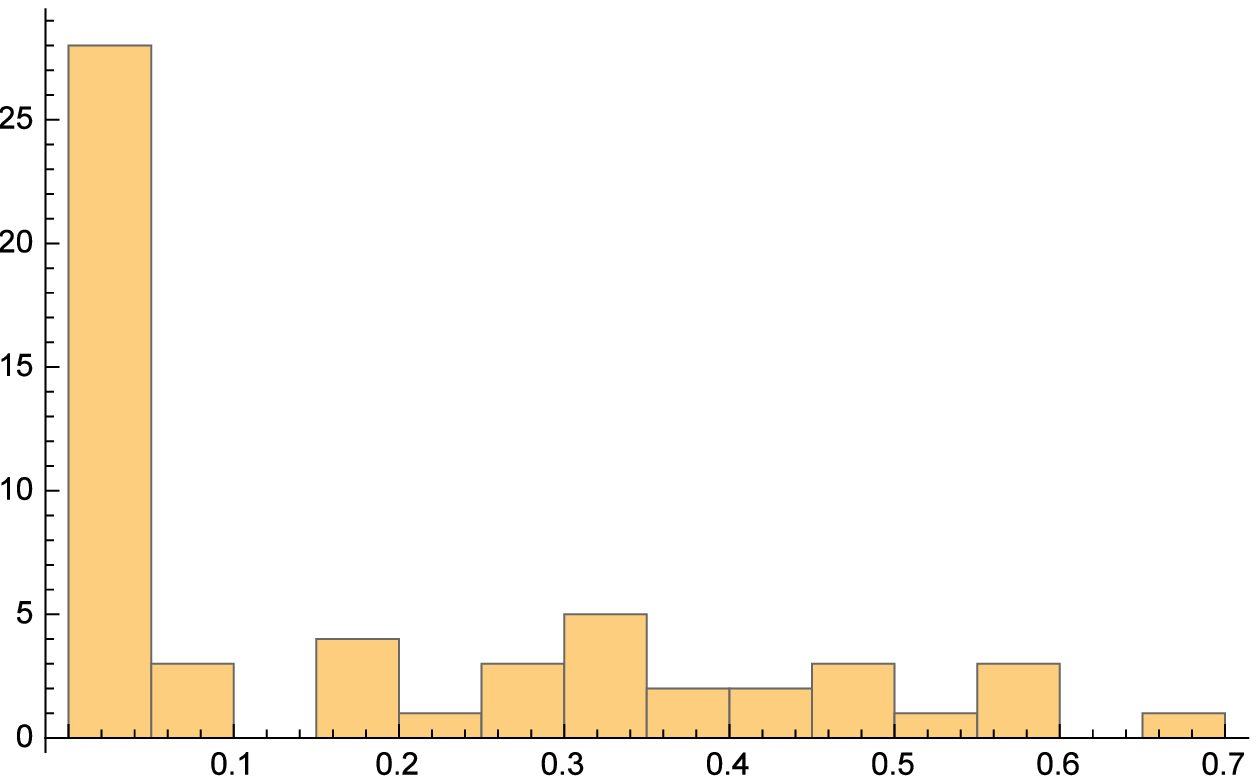} \\
\includegraphics[width=2.5in]{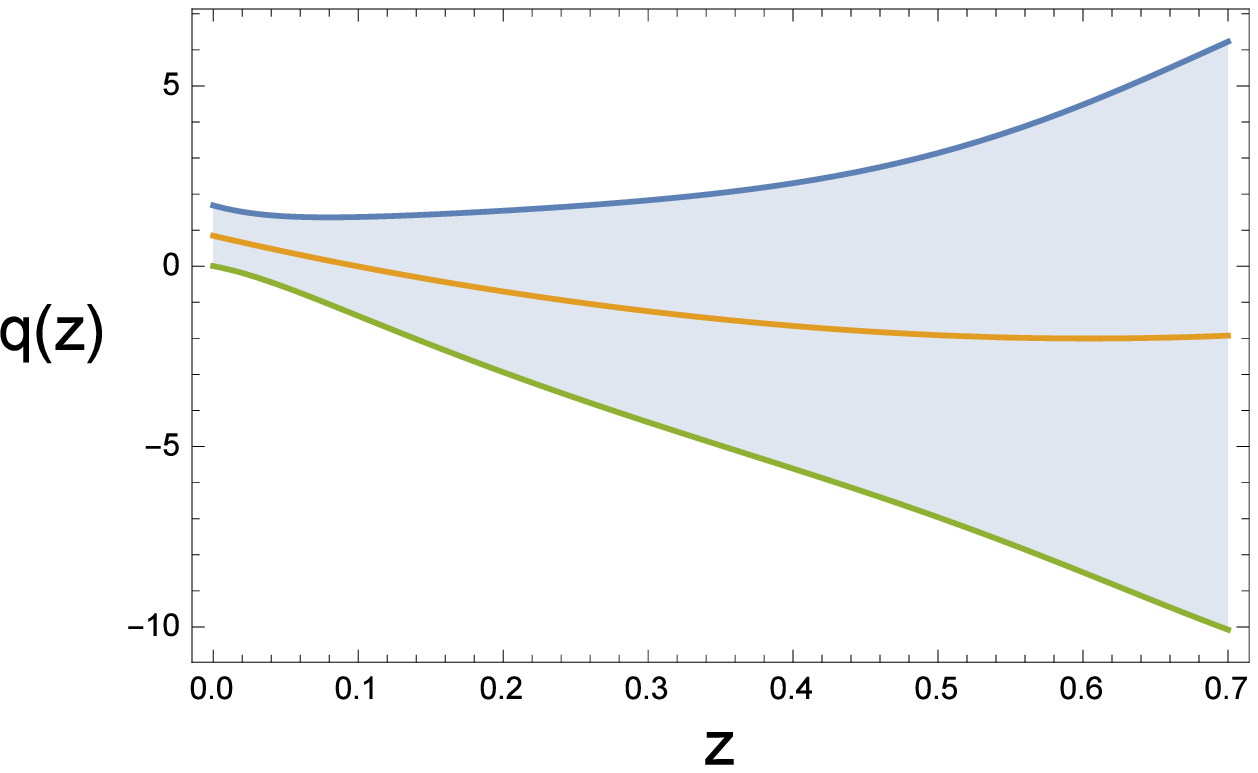}
\end{array}$
\end{center}
\caption{\label{fig: fig11} Redshift distribution for 56 SNIa around
the minimum acceleration region, and the reconstructed deceleration
parameter. The shaded region shows the $1\sigma$ error propagation.}
\end{figure}

Towards the minimum acceleration region, we select 56 SNIa points
with the redshift distribution shown in Fig. \ref{fig: fig11}.
Notice the lack of data for redshift larger than 0.655. The best fit
values for our model gives, $\Omega_m=0.07 \pm 0.61$, $w_0=0.25 \pm
0.59$, and $w_1=-6.7 \pm 8.4$.

The reconstructed deceleration parameter with error propagation at
one sigma is also shown in Fig. \ref{fig: fig11}. Although the best
fit curve transit from an accelerated phase to a decelerated one
around $z \simeq 0.1$, the reconstruction function does not shown
any transition at one sigma. Actually, apparently there is no signal
for the well known decceleration/acceleration transition occurring
around $z \simeq 0.6$.

Given that more than a half of the data (33 points) have redshifts
$z<0.1$, they certainly dominate the parameter estimation process,
leading to this opposite transition (a $q(z)$ changing from $q<0$
for $z>0.1$ to $q>0$ for $z<0.1$). Although this feature is
statistically negligible in this particular case, the increment of
the size of the region and with that, the number of data, the
feature seems to persist as we will see, indicating a less negative
$q$ for $z<0.05$ than for larger redshifts.
\begin{figure}[h!]
\begin{center}$
\begin{array}{c}
\includegraphics[width=2.0in]{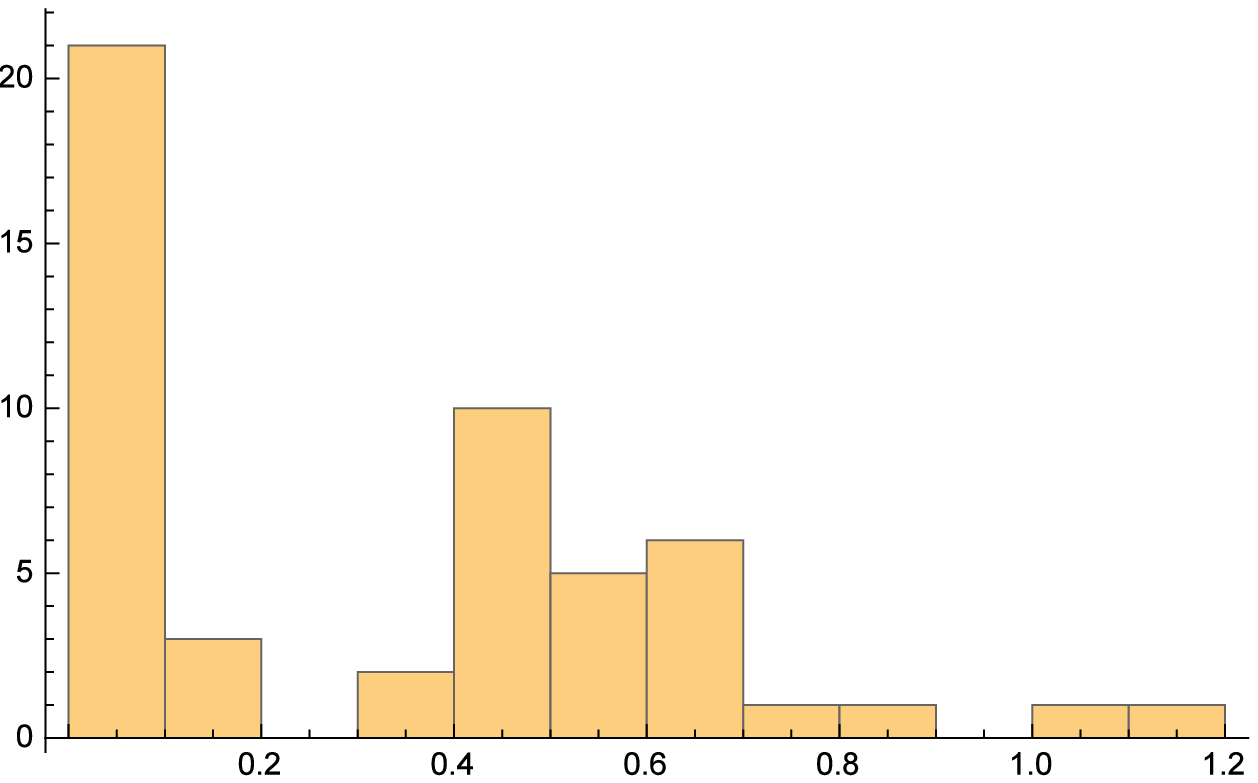} \\
\includegraphics[width=2.5in]{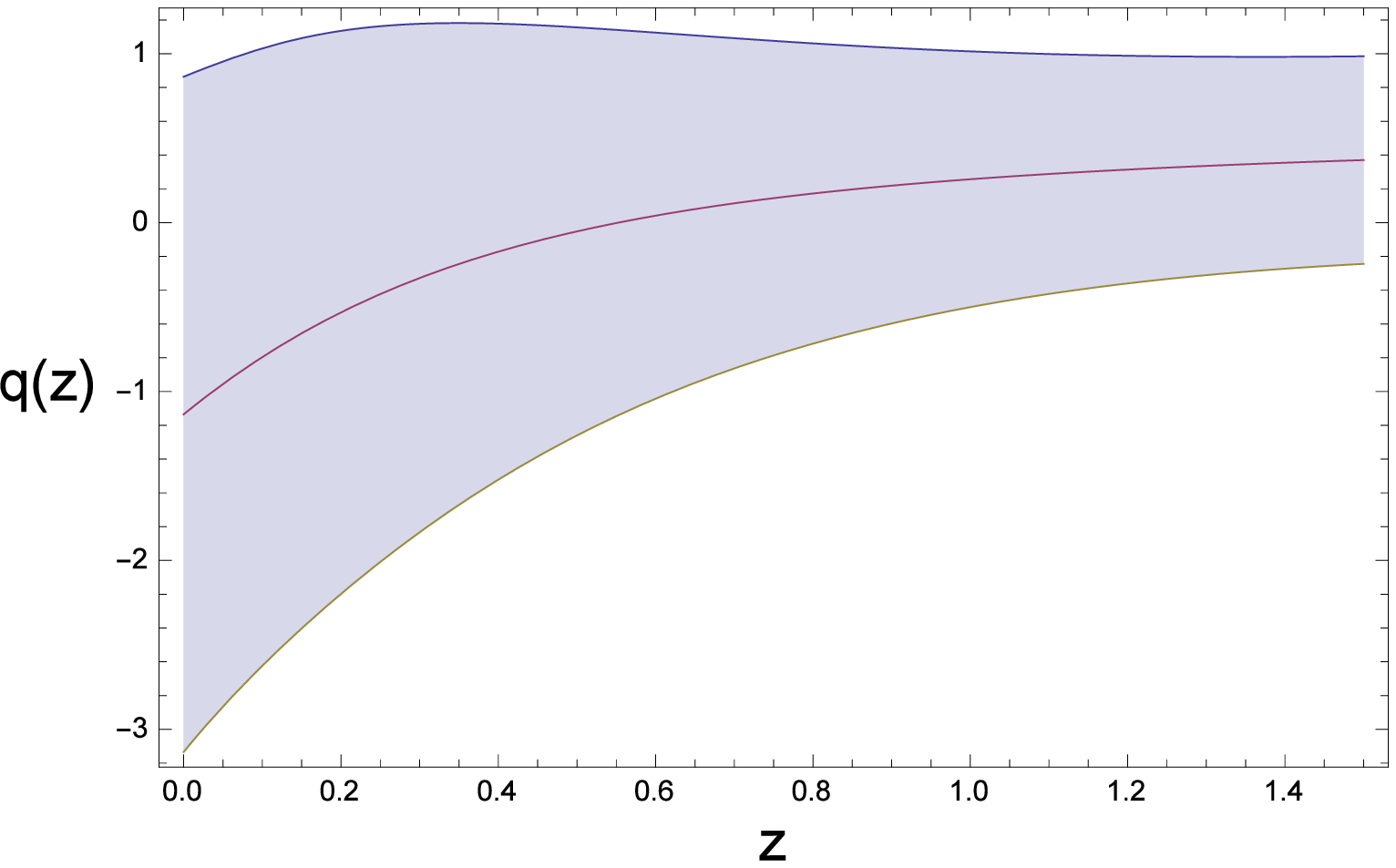}
\end{array}$
\end{center}
\caption{\label{fig: fig12} Redshift distribution for 51 SNIa around
the maximum acceleration region, and the reconstructed deceleration
parameter. The shaded region shows the $1\sigma$ error propagation.}
\end{figure}
In the direction around the maximum acceleration, we select 51 SNIa
data points. The redshift distribution is shown in Fig. \ref{fig:
fig12}.


Most of the data points are concentrated for $z<0.1$, but this time
there are 25 points around $z \simeq 0.55$ and two SNIa with
redshifts higher than 1. The best fit values for our model gives,
$\Omega_m=0.25 \pm 0.57$, $w_0=-1.4 \pm 1.4$, and $w_1=1.8 \pm 4.2$.
The reconstructed deceleration parameter is also shown in Fig.
\ref{fig: fig12}.
Here is clear that the best fit is consistent with a
deceleration/acceleration transition around $z \simeq 0.8$, in
agreement with $\Lambda$CDM. However, as is also the case around the
minimum acceleration region, the trend disappear at $1 \sigma$.

As a matter of conclusion, the use of nearly 50 data points per
region, seems to be not enough to infer any clear trend. However,
the best fit curve obtained in both cases, turns out to be in
agreement with the behavior once more data is added.

\subsection{Region B}

Lets try now regions containing nearly 100 data points each. This
particular case is of interest also in comparison with the results
from the analysis with hemispheres. In fact, here we have about 100
data points around the maximum acceleration, as was also the case in
section III, but the difference here is that around the minimum
region we are taking only 100 points instead of the more than 400
used in that case.

\begin{figure}[h!]
\begin{center}$
\begin{array}{c}
\includegraphics[width=2.0in]{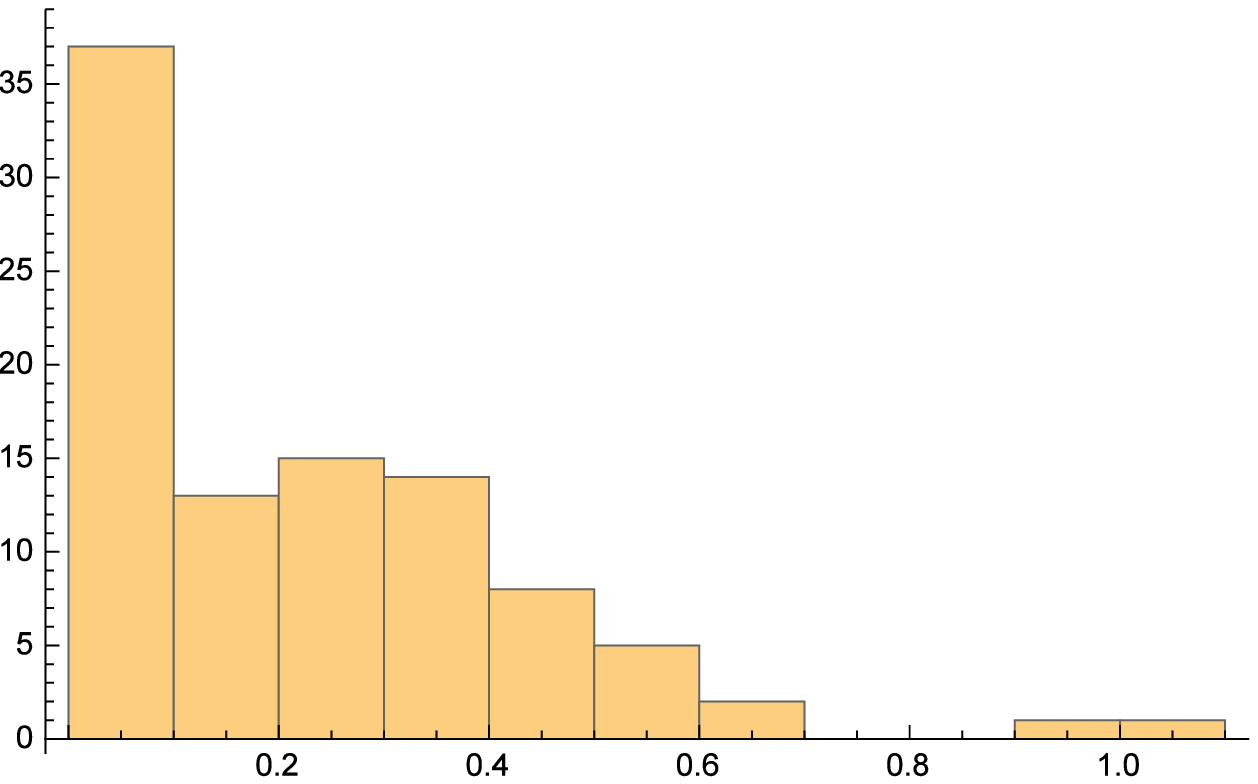} \\
\includegraphics[width=2.5in]{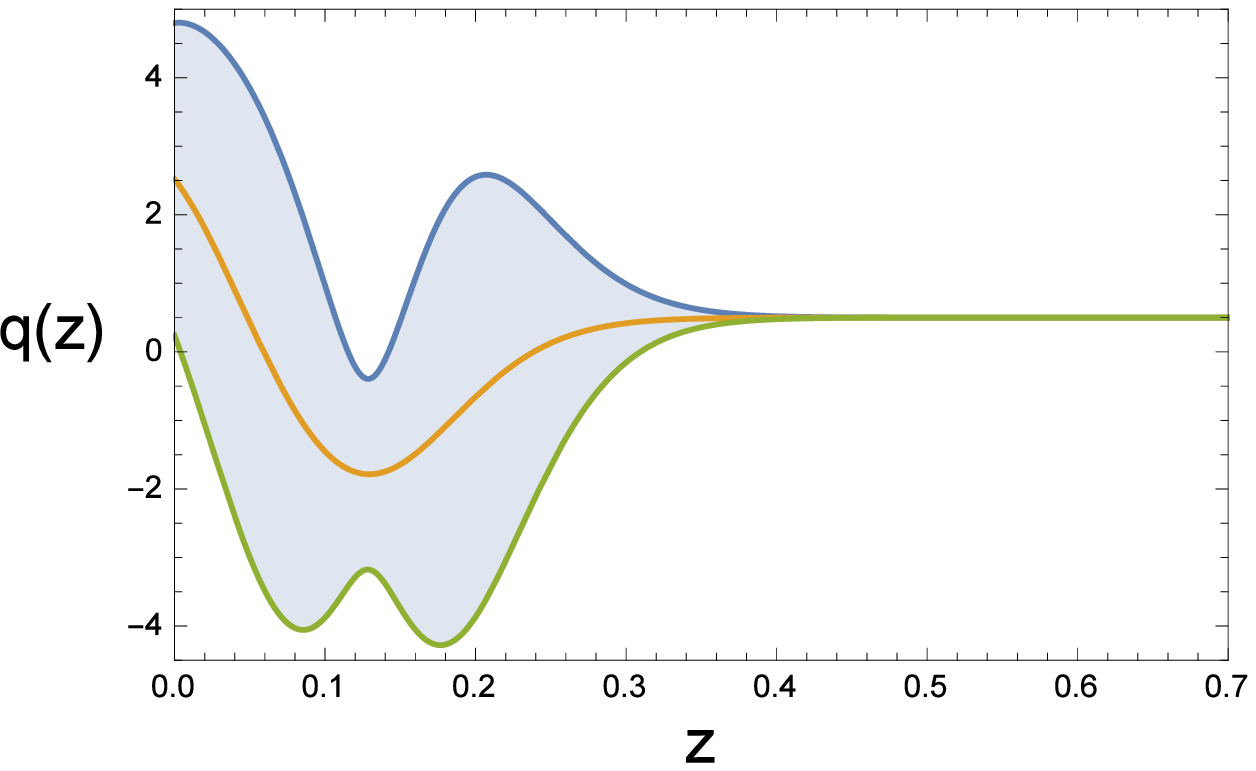}
\end{array}$
\end{center}
\caption{\label{fig: fig13} Redshift distribution for 96 SNIa around
the minimum acceleration region, and the reconstructed deceleration
parameter. The shaded region shows the $1\sigma$ error propagation.}
\end{figure}

Towards the minimum acceleration region, we select 96 SNIa points
with the redshift distribution shown in Fig. \ref{fig: fig13}.
Notice the redshift distribution is now better than those with 50
data points (see Fig.\ref{fig: fig11}). Now the range between
$0.1<z<0.7$ has 57 points well distributed, although still there is
a gap in the range $0.7<z<0.9$. The best fit values for our model
gives, $\Omega_m=0.66 \pm 0.01$, $w_0=3.9 \pm 4.3$, and $w_1=-85 \pm
77$.

The reconstructed deceleration parameter with error propagation at
one sigma is also shown in Fig. \ref{fig: fig13}. In this case the
best fit curve shows a sharper transition at low redshift than the
global one (see the reconstructed $q(z)$ of Fig. (1) obtained with
all data that peaked down to $\simeq -0.5$, meanwhile here the
reconstructed peaked down to $\simeq -2$!!). As we mentioned before,
the trend of $q(z)$ increasing from $z\simeq 0.1$ to $z =0$ that
appears in the case using only 50 data points, here emerges
strongly, even at $1\sigma$.

In the direction around the maximum acceleration, we select 101 SNIa
data points. The redshift distribution is shown in Fig. \ref{fig:
fig14}.
\begin{figure}[h!]
\begin{center}$
\begin{array}{c}
\includegraphics[width=2.0in]{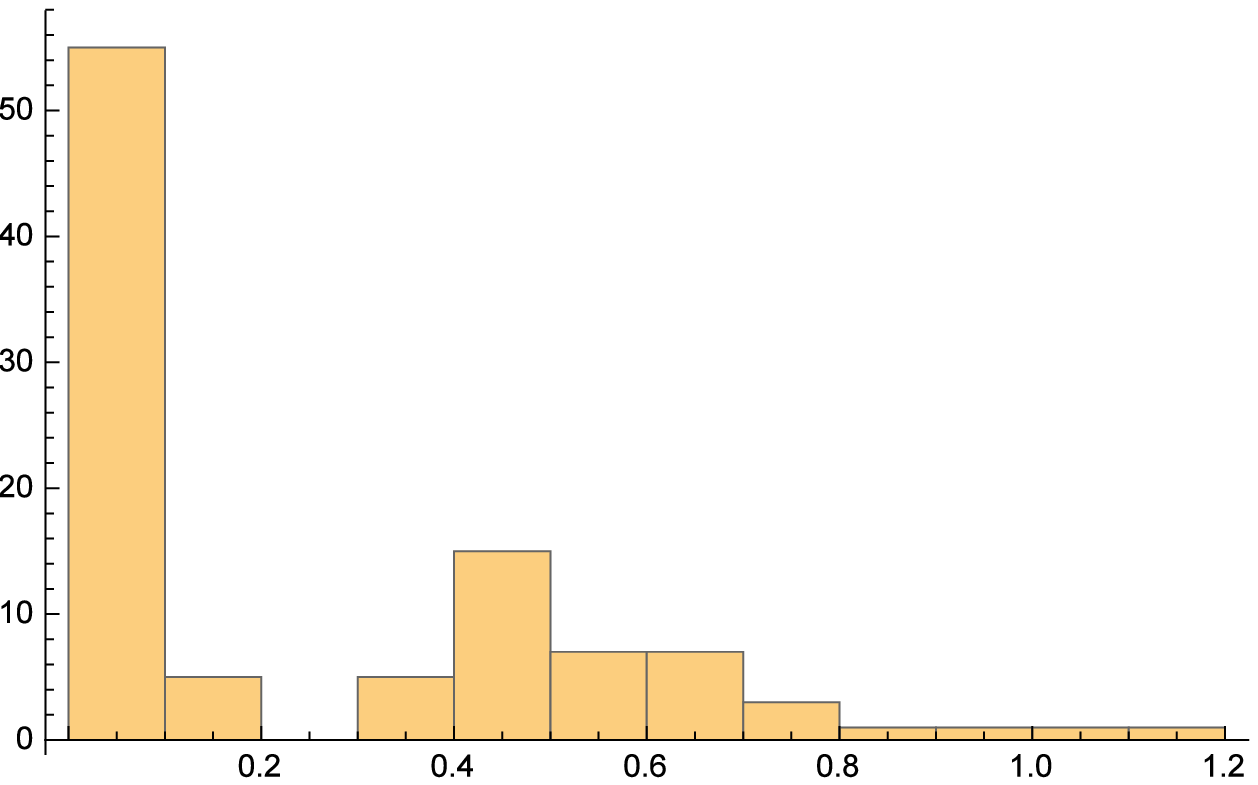} \\
\includegraphics[width=2.5in]{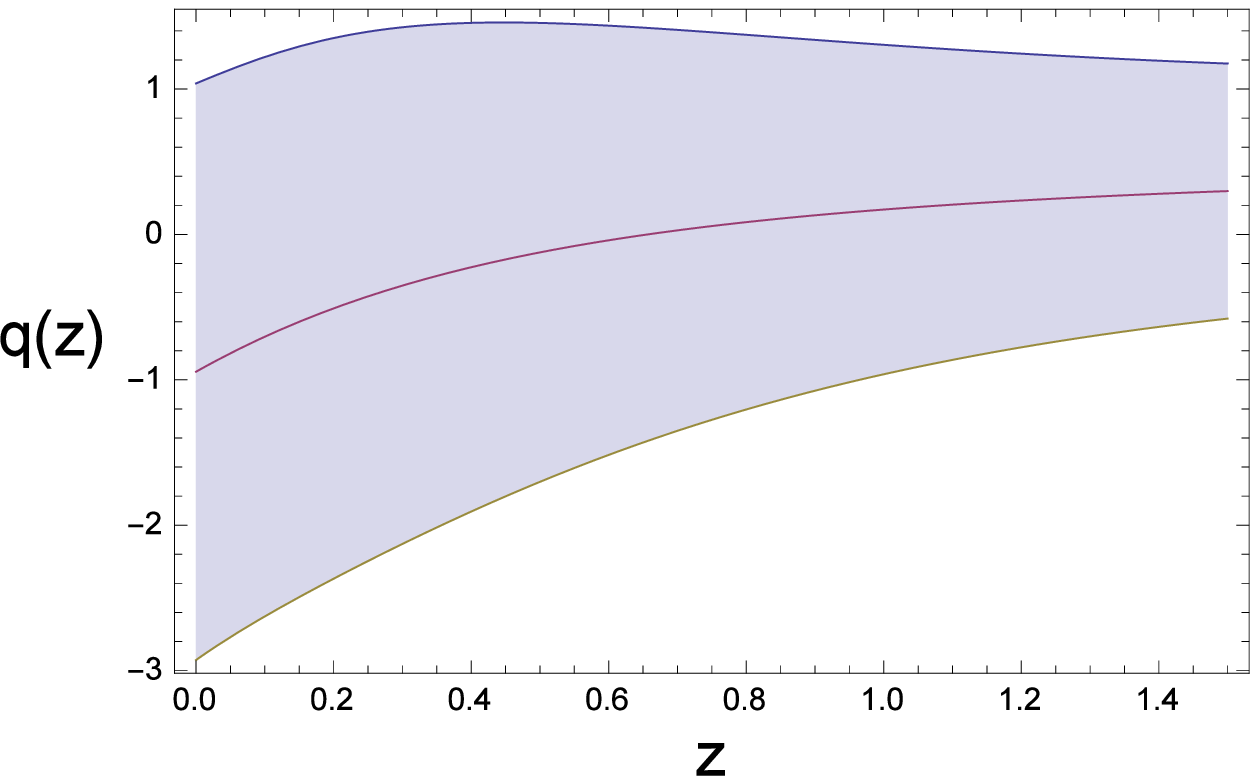}
\end{array}$
\end{center}
\caption{\label{fig: fig14} Redshift distribution for 101 SNIa
around the maximum acceleration region, and the reconstructed
deceleration parameter. The shaded region shows the $1\sigma$ error
propagation.}
\end{figure}
Most of the data points are still concentrated for $z<0.2$, but this
time there are 37 points around $z \simeq 0.55$. The best fit values
for our model gives, $\Omega_m=0.20 \pm 0.72$, $w_0=-1.2 \pm 1.2$,
and $w_1=1.4 \pm 3.3$.

The reconstructed deceleration parameter is also shown in Fig.
\ref{fig: fig14}.
Here is clear that the best fit is again consistent with a
deceleration/acceleration transition around $z \simeq 0.8$, in
agreement with $\Lambda$CDM. However, as is also the case using 51
data points, the trend disappear at $1 \sigma$ (see Fig. \ref{fig:
fig12}).

The use of nearly 100 data points per region is now enough to make a
clear difference between the behavior of the reconstructed $q(z)$:
towards the minimum acceleration region, the low redshift transition
is clearly visible even at one sigma, but in the opposite direction
the trend is completely different, being marginally consistent with
$\Lambda$CDM. We have to stress here that a low redshift transition
of $q(z)$, a bump in the region $0<z<0.2$, occurs in a region with a
robust number of data points (see Fig. \ref{fig: fig13}).


\subsection{Region C}

\begin{figure}[h!]
\begin{center}$
\begin{array}{c}
\includegraphics[width=2.0in]{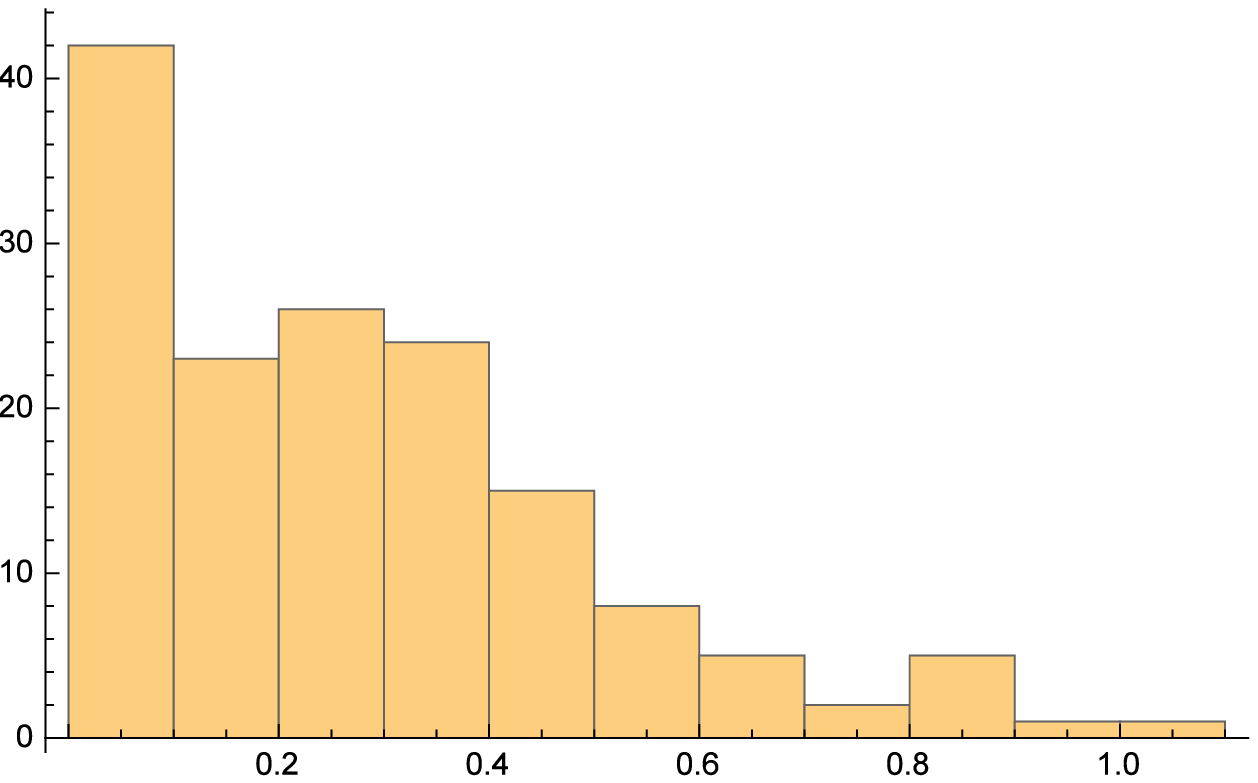} \\
\includegraphics[width=2.5in]{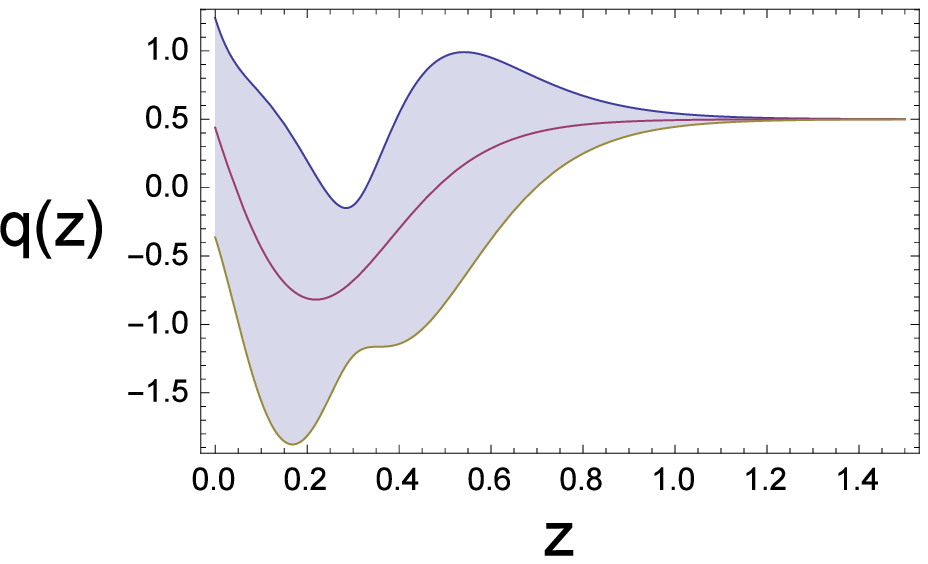}
\end{array}$
\end{center}
\caption{\label{fig: fig15} Redshift distribution for 152 SNIa
around the minimum acceleration region, and the reconstructed
deceleration parameter. The shaded region shows the $1\sigma$ error
propagation.}
\end{figure}

In this region we select around 150 SNIa per region. In the one
pointing to the minimum acceleration we get 152 data points, from
which we obtain the best fit shown in Table \ref{tblcdm}. The
redshift distribution is shown in Fig.\ref{fig: fig15}.
The reconstructed deceleration parameter is also shown in that
figure.
\begin{table}[h!]
\begin{scriptsize}
\begin{center}
\begin{tabular}{cl ccc}
\hline
   &     $\chi^2_{min}/dof$  &         $\Omega_{m}$        &$ \omega_0$      &$ \omega_1 $\\
\hline
   Region C \cr
1)   &   $124.03/133$ &  $0.17 \pm 1.24$  &  $-1.1 \pm 1.7 $  & $1.2 \pm 3.8$  \\
2)   &   $134.10/149$  &  $0.44 \pm 0.13 $ &  $-0.07\pm 0.96$  & $-12\pm 14$\\

\hline
   Region D \cr
1)   &   $190.72/194 $  & $0.40 \pm 0.08 $ &  $-0.64\pm 0.77$  & $-7.9\pm 9.6$\\
2)   &   $185.284/196$ & $0.47 \pm 0.11 $  &  $-0.1 \pm 1.1 $ & $-14 \pm 16$\\
\hline

\end{tabular}
\end{center}
\end{scriptsize}
\caption[Constraints on $CPL$ model]{Constraints on the cosmological
parameters of $CPL$ model for the two levels. Region C take
approximately 150 SNIa points, and Region D approximately 200 SNIa
points. The number 1) is for direction of maximum acceleration, 2)
is for direction of minimum acceleration.\label{tblcdm}}
\end{table}

\begin{figure}[h!]
\begin{center}$
\begin{array}{c}
\includegraphics[width=2.0in]{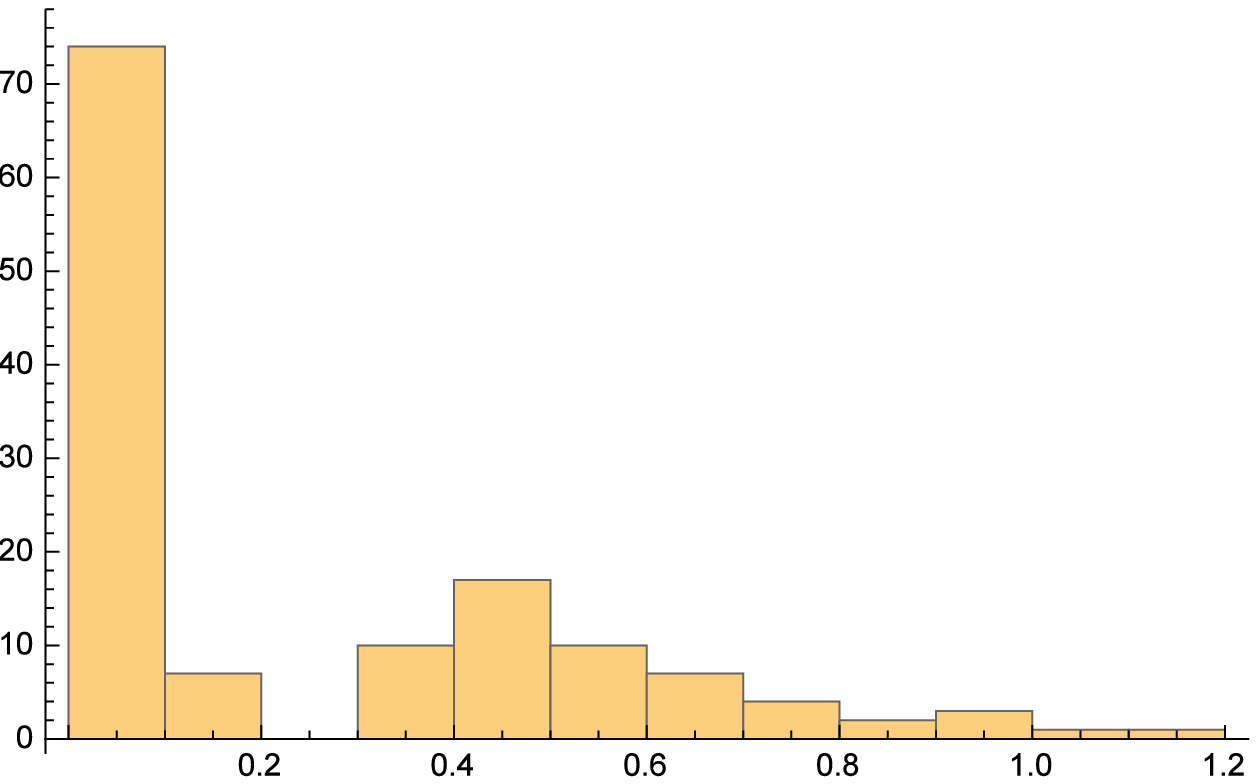} \\
\includegraphics[width=2.5in]{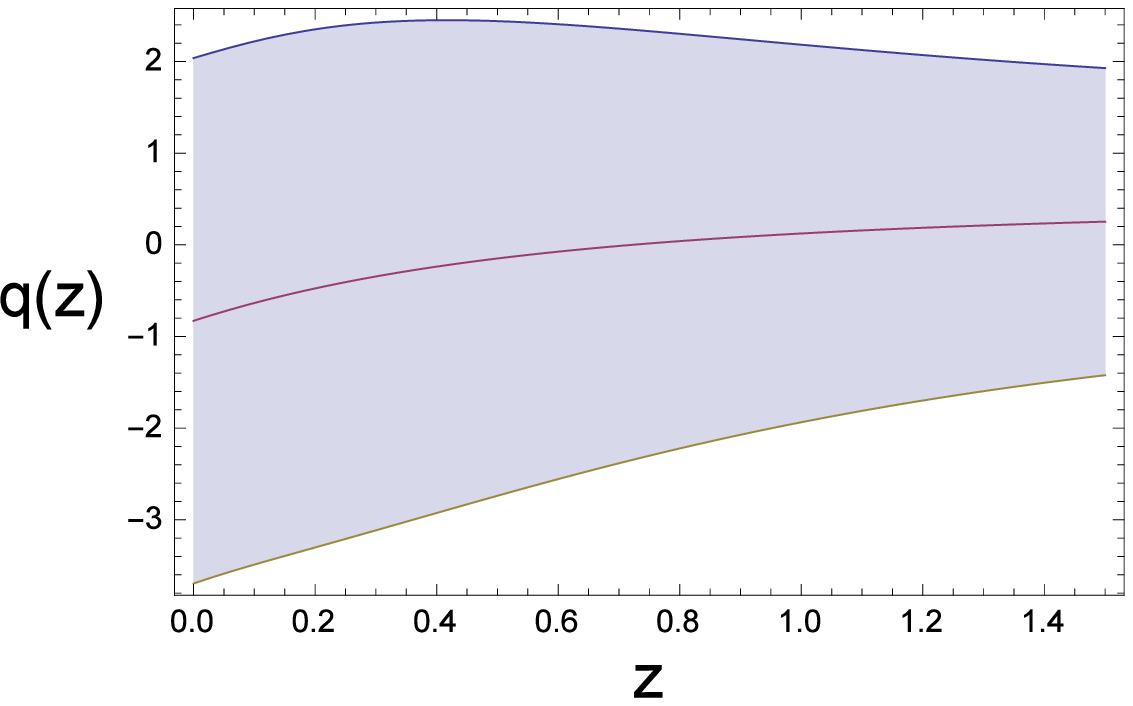}
\end{array}$
\end{center}
\caption{\label{fig: fig16} Redshift distribution for 136 SNIa
around the maximum acceleration region, and the reconstructed
deceleration parameter. The shaded region shows the $1\sigma$ error
propagation.}
\end{figure}

The same is done for the opposite direction, that pointing towards
the maximum acceleration. In this case we select 136 data points.
The redshift distribution is shown in Fig. \ref{fig: fig16}

As we notice looking at Fig. \ref{fig: fig15} and \ref{fig: fig16},
using 150 data points we obtain essentially the same results as in
the previous section (using around 100 points) and those obtained in
section III using the hemispheres.
However it is interesting to notice that, as we have mentioned
before, from Fig. \ref{fig: fig15} it is clear that at $1\sigma$ the
deceleration/acceleration transition occurring around $z\simeq 0.5$
is clearly visible in the reconstructed $q(z)$, at the same level as
is also visible the low-redshift transition (around $z\simeq 0.2$).
In the opposite direction, from Fig. \ref{fig: fig16} at $1\sigma$
is not possible to talk about a deceleration/acceleration transition
of $q(z)$.

It is interesting to investigate what happens once we add more data
points per region. We select around 200 data points for region D.
Our results show that the behavior found in that case persist once
more data is added.

\subsection{Region D}

\begin{figure}[h!]
\begin{center}$
\begin{array}{c}
\includegraphics[width=2.0in]{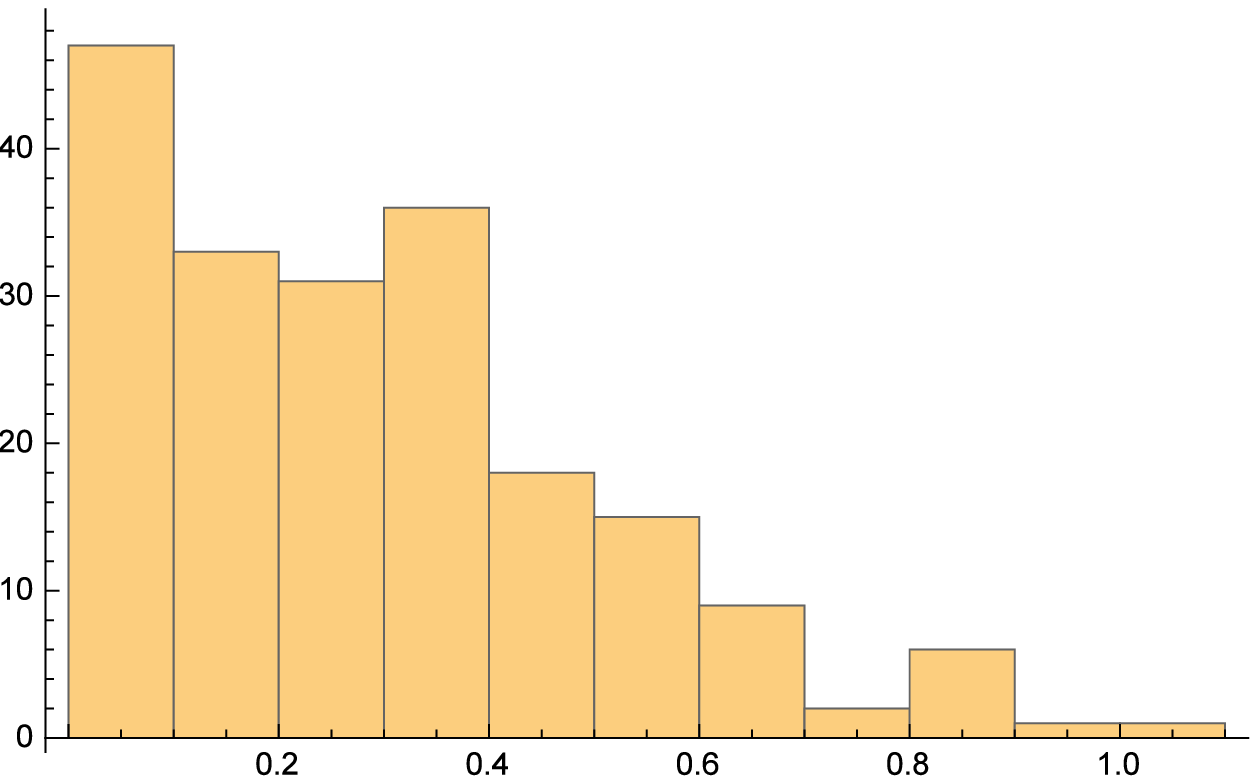} \\
\includegraphics[width=2.5in]{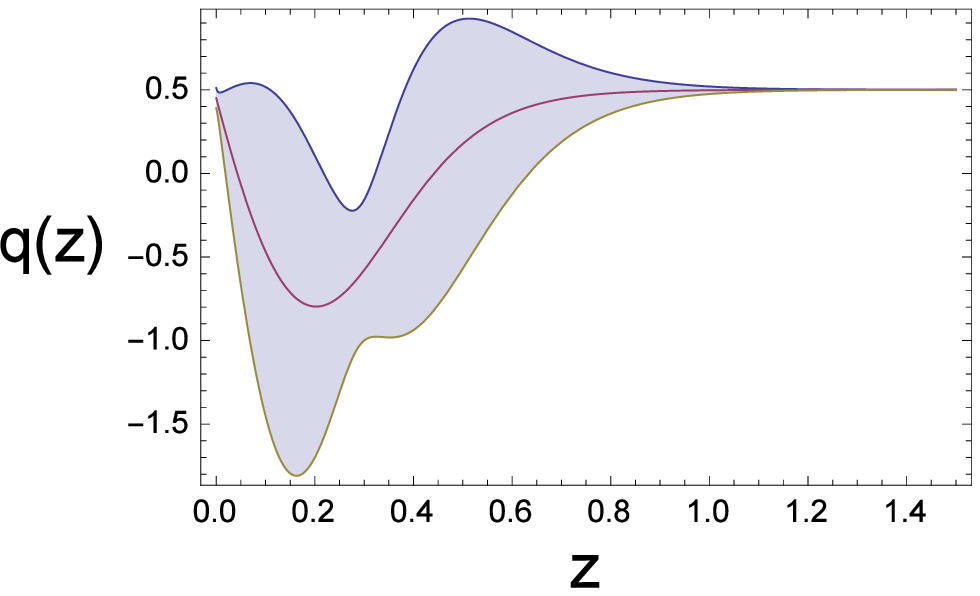}
\end{array}$
\end{center}
\caption{\label{fig: fig17} Redshift distribution for 199 SNIa
around the minimum acceleration region, and the reconstructed
deceleration parameter. The shaded region shows the $1\sigma$ error
propagation.}
\end{figure}

\begin{figure}[h!]
\begin{center}$
\begin{array}{c}
\includegraphics[width=2.0in]{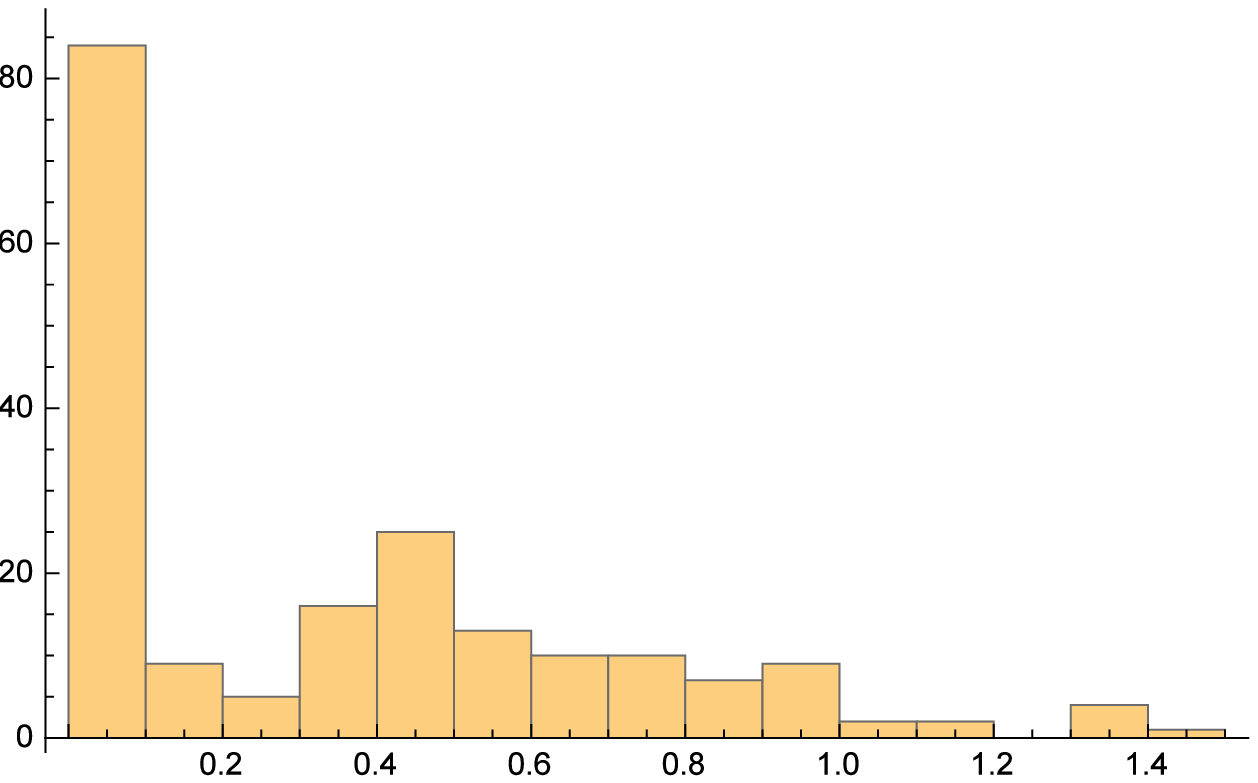} \\
\includegraphics[width=2.5in]{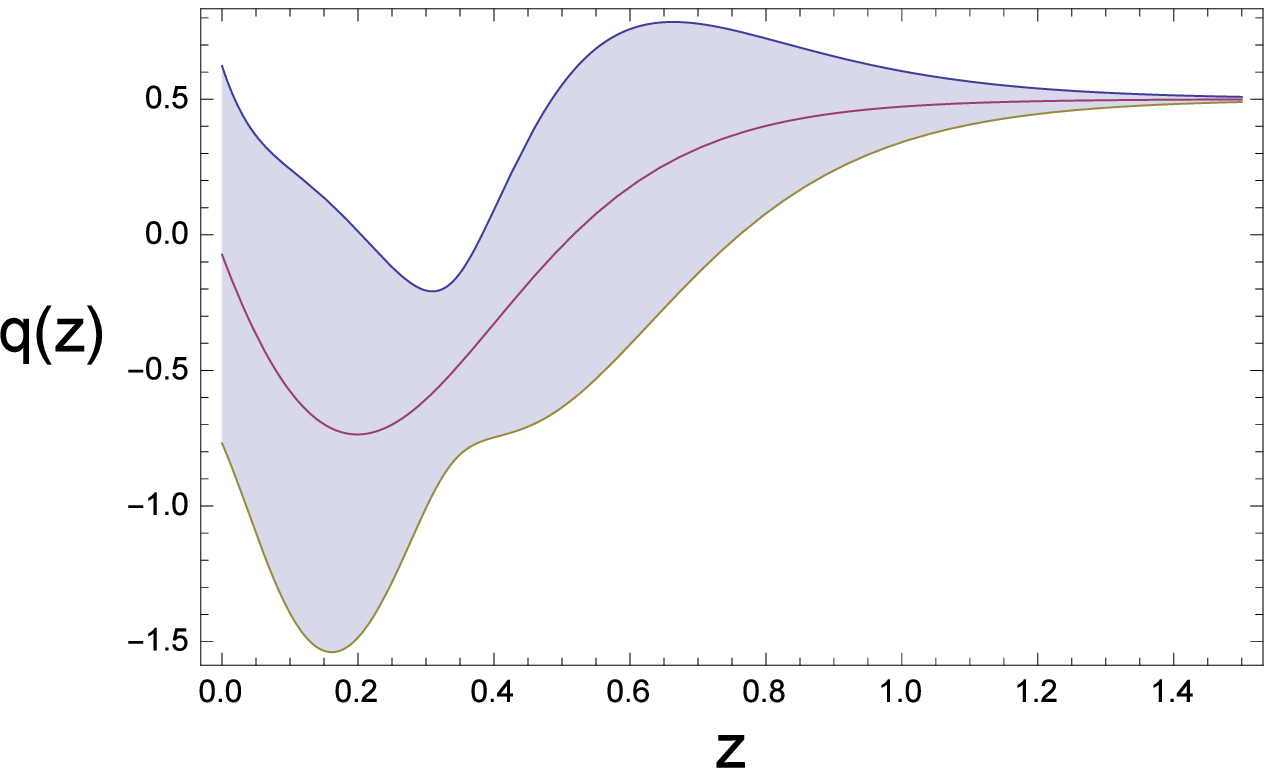}
\end{array}$
\end{center}
\caption{\label{fig: fig18} Redshift distribution for 197 SNIa
around the minimum acceleration region, and the reconstructed
deceleration parameter. The shaded region shows the $1\sigma$ error
propagation.}
\end{figure}

In the region pointing towards the minimum acceleration we select
199 SNIa, and in the opposite direction (towards maximum
acceleration) we select 197 SNIa. The redshift distribution and the
reconstructed deceleration parameter in the first case are displayed
in Fig. \ref{fig: fig17}
meanwhile in the case around the maximum acceleration region are
displayed in Fig. \ref{fig: fig18}.


The results show that adding these 61 data points (in the region
around the maximum acceleration) is enough to change the best fit
and obtain something similar to the global fit shown in
Fig.\ref{fig:fig1}. This result should not be a surprise, because
when adding more data points we are increasing the size of the
region around the maximum acceleration point. Remember that using
the hemispheres, the number of data points on the side of maximum
acceleration was only 116.



The result of the analysis for the regions with 100 and 150 data
points are consistent with our previous result, showing a $q(z)$
consistent with LCDM around the maximum acceleration point, and a
low redshift transition for $q(z)$ in the opposite direction. The
analysis of the region with 200 data points converge -- in both
directions -- to the global result presented in section II, that
showing a low redshift transition for $q(z)$. This is not surprising
because the number of SNIa data points are more numerous around the
minimum acceleration point.

\section{Analysis using the Loss sample}

We have also performed the study of the consequences, in the
reconstructed deceleration parameter, of the existence of an axis of
maximal asymmetry using the Loss data set
\cite{Ganeshalingam:2013mia}. In order to find this axis, we follow
the procedure of \cite{antoniou}, using a flat $\Lambda$CDM as a
base model, and start a scan of random axes looking for the maximum
variation in the best fit of $\Omega_m$. Using the Loss sample we
found that the direction of maximal acceleration points towards
$(l,b)=(309^o,31^o)$.

In the hemisphere corresponding to the maximum acceleration, there
are 216 supernovae and in the other the remaining 370. Again we
observe the asymmetry in the distribution, although less severe than
in the case of the Union 2 sample. In fact, using the Union 2 set we
have verified the value $(\Delta
\Omega_{m})_{\textrm{max}}/\Omega_m=0.43 \pm 0.06$ from
\cite{antoniou}. Using the same procedure with the Loss sample we
get $(\Delta \Omega_{m})_{\textrm{max}}/\Omega_m=0.30 \pm 0.06$.

The best fit values in each hemisphere are displayed in Table
\ref{tab:table02}:

\begin{table}[h!]
\begin{center}
\begin{tabular}{ccccc}
\hline
\\
 & $\chi^2_{min}/dof$ & $\Omega_{m}$ & $\omega_0$ & $\omega_1 $ \\

\hline
A   &   $197.283/213$ & $0.16 \pm 1.08 $ & $ -0.9 \pm 0.9 $ & $0.2 \pm 5.6$  \\
B   &   $374.76/367$ & $0.39 \pm 0.07$ & $-0.47 \pm 0.43 $ &  $-7.6 \pm 6.2$\\
\hline
\end{tabular}
\end{center}
\caption{Best fit values of the cosmological parameters using the
LOSS sample separating the sky in two hemispheres. Row A for the 216
SNIa in the maximum acceleration hemisphere. Row B the 370 SNIa in
the opposite direction. See also Fig.(\ref{fig01}).
\label{tab:table02}}
\end{table}

\begin{figure}[h!]
\centering \leavevmode\epsfysize=9.0cm \epsfbox{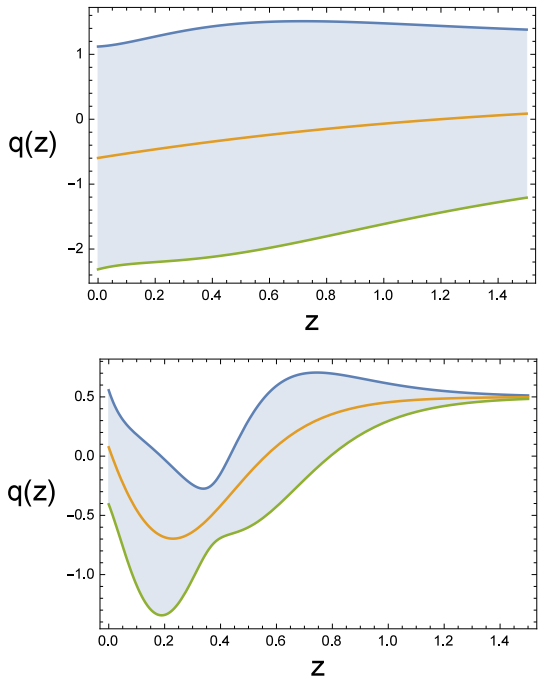}
\caption{ Reconstructed deceleration parameter $q(z)$ using data
from two hemispheres using the Loss sample. The panels show the one
around the maximum (upper) and minimum (lower) acceleration points
respectively.
See Table
\ref{tab:table02} for the best fit values for each case. }
\label{fig01}
\end{figure}

In agreement with the previous analysis using the Union 2 data set,
the result using the hemisphere towards the maximum acceleration is
consistent with $\Lambda$CDM, showing no more than a $0.1\sigma$
departure from it (see Fig. \ref{fig01}).
Our results are also in agreement with the previous case, the
analysis with data from the hemisphere of minimum acceleration shows
a sharp low redshift transition. In this case the best fit is around
$1.5\sigma$ away from the $\Lambda$CDM value.

It would be interesting to study this effect in the newest data set,
as the JLA set \cite{betoule} comprising more than 740 data points,
and also using different parameterizations, as we have performed in
\cite{Magana:2014voa}. These issues are under study.

\section{Discussion}

In this paper we have study the effects in the reconstructed
deceleration parameter using the well known dipolar asymmetry
present in SNIa data. We have used the Union 2 set and the result of
\cite{antoniou}, a dipolar asymmetry with a maximum acceleration
towards $(l,b)=(309^o,18^o)$ and a minimum acceleration towards
$(l,b)=(129^o,-18^o)$ is found. The low reshift transition,
previously recognized in \cite{shafi2009, Li:2010da, Cardenas:2011a,
Cardenas:2013roa, cardenas:2014, Magana:2014voa}, appears only
towards the direction of minimum acceleration, and it is absent
towards the opposite direction. This result was obtained separating
the data in two hemispheres constructed along the maximal asymmetry
axis.

Using an approximately equal number of data points around each
direction, we study four regions with 50, 100, 150 and 200 data.
From this analysis we found hints of a low-redshift transition
towards the minimum acceleration region that, apparently given the
amount of data in that region, turns out to be dominant once all the
sky is taken into account. Once this low redshift transition arise,
in the opposite direction the reconstructed $q(z)$ seems featureless
showing no apparent evolution at $1\sigma$, however being consistent
with $\Lambda$CDM.

We have also performed a similar analysis using the Loss dataset
\cite{Ganeshalingam:2013mia} finding the results displayed in Fig.
\ref{fig01}, which are in agreement with those using the Union 2
dataset (see Fig. \ref{fig00}).

Assuming the result of \cite{Vanderveld:2006rb} as our main
hypothesis -- i.e., a low redshift transition of $q(z)$ can be
considered as a signal of a local under-density --
 our work indicates that the supernova data is able to detect such an
under-density only towards one direction in the sky. This result
suggest that a better description of the data would be a background
metric with a dipolar asymmetry.

There remains to check if this asymmetry in the behavior of the
reconstructed deceleration parameter is not due to the anisotropic
distribution of data. This issue as been discussed in the past. As
we have mentioned before, using the Union 2 dataset, the authors of
\cite{antoniou} found a preferred axis in the data, but they also
checked if such anisotropy can be obtained using simulated isotropic
data. Their method suggest that only a $30 \%$ of the simulated data
can reach such anisotropy, implying that the anisotropy of the Union
2 set is consistent with statistical isotropy. However, the
coincidence of the axes (a total of six phenomenological axes
identified in \cite{antoniou}) within a small angular region in the
sky, makes this dipolar anisotropy a feature that needs further
scrutiny. Using the Loss sample, the authors of
\cite{Appleby:2014kea} found a hint for anisotropy -- most
pronounced in the range $0.015<z<0.045$ -- but once they take into
account large scale velocity perturbations, the results shows no
evidence for any anomalous deviation from the isotropic
$\Lambda$CDM. In \cite{Bengaly:2015dza} the authors studied this
issue using the Union 2.1 and JLA set. Although they found that both
datasets are statistically anisotropic, the correlation with the
anisotropic distribution found turns out to be different: using the
Union 2.1 they found a small correlation but using the JLA sample
they found a stronger correlation. A similar study
\cite{Huterer:2015gpa} using the JLA dataset also showed that by
ignoring the velocity covariance may produce a hint of anisotropy
from the data. In this context, it is interesting to notice here the
result of the work \cite{Nielsen:2015pga} where the authors studied
the JLA set, considering all the information from possible
systematics (encoded in the covariance) in the analysis, finding a
marginal (less than $3\sigma$) evidence for the accepted cosmic
acceleration.

Altogether, the apparent detection of a local under density can be
considered as a possible cause of the original tension between low
and high redshift observational constraints discussed in
\cite{shafi2009, Li:2010da, Cardenas:2011a, Cardenas:2013roa,
Magana:2014voa}, which as a solution also agree with the analysis of
\cite{Keenan:2013mfa} where they proposed the existence of a local
under-density to alleviate the tension in the determination of
$H_0$.

Further study on this subject would be interesting. For example, the
effects of using different parameterizations, and also the use of
different SNIa data sets. Given that the SNIa data used in this work
is obtained after the calibration of the whole set (by fitting
globally also the $\alpha$, $\beta$ and $M$ parameters, being the
last one connected with the value of $H_0$) it would be interesting
to study the impact of the hemispherical asymmetry on such a
calibration, and then on the cosmological parameters.
All these issues are under study.

\section*{Acknowledgments}

The authors want to thank Juan Maga\~na for useful discussions. VHC
acknowledges financial support through DIUV project No. 13/2009, and
FONDECYT 1110230. C.B. and V.M. acknowledges financial support
through FONDECYT 1120741. V.M. acknowledges the support from Centro
de Astrof\'{\i}sica de Valpara\'{\i}so.


\end{document}